%% file: main.tex
\documentclass[9pt]{article}
\def\pagenum{1}

\pdfoutput=1

\input{files/configs}

\usepackage{natbib}
\setcitestyle{numbers,square}
\usepackage{graphicx}
\usepackage{float}
\usepackage{indentfirst} 
\usepackage{bigstrut,multirow,rotating,booktabs}
\usepackage{lineno}
\usepackage{titlesec}
\usepackage{hyperref} 
\usepackage{cleveref} 
\usepackage{tabularx}

\hypersetup{colorlinks = true, 
	    linkcolor = blue, 
	    urlcolor = blue,
        citecolor = blue} 

\usepackage{etoolbox}
\BeforeBeginEnvironment{tabular}{\footnotesize}

\usepackage{enumitem}
\setitemize[1]{itemsep=0pt,partopsep=0pt,parsep=\parskip,topsep=5pt}

\usepackage{authblk}

\crefname{table}{Table}{Table}
\crefname{figure}{Figure}{Figure}

\def\thetitle{{A neuroergonomics model to evaluating nuclear power plants operators' performance under heat stress driven by ECG time-frequency spectrums and fNIRS prefrontal cortex network: a CNN-GAT fusion model.}} 
\def\authorOne{\authorfont{Yan Zhang}}
\def\authorTwo{\authorfont{Ming Jia}}
\def\authorThree{\authorfont{Meng Li}}
\def\authorFour{\authorfont{JianYu Wang}}
\def\authorFive{\authorfont{XiangMin Hu}}
\def\authorSix{\authorfont{ZhiHui Xu}}
\def\authorSeven{\authorfont{Tao Chen}}

\def\institutionOne{\subauthorfont{State Key Laboratory of Nuclear Power Safety Technology and Equipment, China Nuclear Power Engineering Co., Ltd., Shenzhen, Guangdong, 518172, China}}
\def\institutionTwo{\subauthorfont{School of Safety Science, Tsinghua University, Beijing 100084,  P.R. China}}
\def\institutionThree{\subauthorfont{Anhui Province Key Laboratory of Human Safety, Hefei Anhui 230601, China}}
\def\institutionFour{\subauthorfont{Beijing Key Laboratory of Comprehensive Emergency Response Science, Beijing, China}}

\begin{document}

\interfootnotelinepenalty=100000

\title{\TitleFont{\thetitle}}
\author[b]{\authorOne}
\author[a]{\authorTwo}
\author[b]{\authorThree}
\author[b]{\authorFour}
\author[b]{\authorFive}
\author[a,$^*$]{\authorSix}
\author[b,c,d,\footnote{Corresponding author. \\ E-mail addresses: xuzhihui@hrbeu.edu.cn (Z.H. Xu), chentao.a@tsinghua.edu.cn (T. Chen).}]{\authorSeven} 

\affil[a]{\institutionOne}
\affil[b]{\institutionTwo}
\affil[c]{\institutionThree}
\affil[d]{\institutionFour}

\vspace{-1mm}
\date{}                                                 
\maketitle                                              
        

\input{files/01abstract}

\thispagestyle{fancy}

\input{files/02introduction}
\input{files/03method}

\input{files/04results}

\input{files/05discussion}
\input{files/06conclusion}
\input{files/07acknowledge}


\bibliographystyle{elsarticle-num}
\bibliography{main.bib}

\end{document}

%% file: files/configs.tex
\usepackage{times}                                     
\usepackage{amsfonts,amsmath,amssymb, amsthm}

\usepackage{latexsym}
\usepackage[a4paper]{geometry}

\usepackage{fancyhdr}
\usepackage{placeins}                           
\usepackage{xcolor}
\usepackage[utf8]{inputenc} 
\usepackage{graphicx}%
\usepackage{tabularx}
\usepackage{array}
\newcolumntype{x}[1]{%
>{\raggedleft\hspace{0pt}}p{#1}}%

\usepackage{booktabs}

\usepackage{comment}

\usepackage[labelfont=bf,labelsep=period, skip=5pt]{caption}

\usepackage{pgf,tikz}
\usepackage{mathrsfs}
\usetikzlibrary{arrows}

\usepackage{pgfplots}
\pgfplotsset{width=7cm,compat=1.10}

\usepackage{caption}
\usepackage{subcaption}
\usepackage{float}

\usepackage{outlines}
\usepackage{enumitem}
\setenumerate[1]{label=\arabic*.}
\setenumerate[2]{label=(\alph*).}
\setenumerate[3]{label=\roman*.}
\setenumerate[4]{label=\alph*.}

\RequirePackage{color}
\definecolor{cej@color}{rgb}{0.5,0.37109375,0.59765625}
\definecolor{abs}{rgb}{0.251,0.541,0.72.9}
\definecolor{ffqqqq}{rgb}{1,0,0}
\definecolor{qqqqff}{rgb}{0,0,1}
\definecolor{cqcqcq}{rgb}{0.75,0.75,0.75}
\definecolor{ttqqqq}{rgb}{0.2,0,0}
\definecolor{ffxfqq}{rgb}{1,0.5,0}
\definecolor{uuuuuu}{rgb}{0.27,0.27,0.27}
\definecolor{ffqqff}{rgb}{1,0,1}
\definecolor{xfqqff}{rgb}{0.5,0,1}
\definecolor{xdxdff}{rgb}{0.49,0.49,1}
\definecolor{red}{rgb}{1,0,0}
\definecolor{blue}{rgb}{0,0,1}
\definecolor{ccqqqq}{rgb}{0.8,0,0}
\definecolor{fftttt}{rgb}{1,0.2,0.2}
\definecolor{qqwuqq}{rgb}{0,0.39,0}
\definecolor{qqqqff}{rgb}{0,0,1}
\definecolor{wrwrwr}{rgb}{0.3803921568627451,0.3803921568627451,0.3803921568627451}
\definecolor{sexdts}{rgb}{0.1803921568627451,0.49019607843137253,0.19607843137254902}
\definecolor{rvwvcq}{rgb}{0.08235294117647059,0.396078431372549,0.7529411764705882}
\definecolor{ccqqqq}{rgb}{0.8,0,0}
\definecolor{qqqqff}{rgb}{0,0,1}
\definecolor{wwwwww}{rgb}{0.4,0.4,0.4}
\definecolor{qqwuqq}{rgb}{0,0.39215686274509803,0}
\definecolor{uuuuuu}{rgb}{0.266,0.266,0.266}

\usepackage{titlesec}
\titleformat{\section}[block]{\normalsize\scshape\bfseries{\arabic{section}.}}{}{1em}{}
\titleformat{\section}[block]{\normalsize\scshape\bfseries}{\thesection}{1em}{}
\titleformat*{\subsection}{\normalsize\scshape\itshape\bfseries}
\titleformat*{\subsubsection}{\small\scshape\itshape}
\usepackage{framed}
\usepackage[colorlinks,allcolors=black]{hyperref}

\newcounter{ggbFirstpage}
\newcommand*{\TitleFont}{%
      \usefont{\encodingdefault}{\rmdefault}{}{n}%
      \fontsize{16}{18}%
      \selectfont}

\def\authorfont{\normalfont\fontsize{12}{14}\selectfont\centering}
\def\subauthorfont{\normalfont\fontsize{10}{12}\selectfont\centering}
\definecolor{wrwrwr}{rgb}{0.3803921568627451,0.3803921568627451,0.3803921568627451}
\definecolor{sexdts}{rgb}{0.1803921568627451,0.49019607843137253,0.19607843137254902}
\definecolor{rvwvcq}{rgb}{0.08235294117647059,0.396078431372549,0.7529411764705882}


%% file: files/01abstract.tex
%
\bgroup
\color{abs}
\hrule
\egroup

%

\begin{abstract}

Operators experience complicated physiological and psychological states when exposed to extreme heat stress, which can impair cognitive function and decrease performance significantly, ultimately leading to severe secondary disasters. Therefore, there is an urgent need for a feasible technique to identify their abnormal states to enhance the reliability of human-cybernetics systems. With the advancement of deep learning in physiological modeling, a model for evaluating operators' performance driven by electrocardiogram (ECG) and functional near-infrared spectroscopy (fNIRS) was proposed, demonstrating high ecological validity. The model fused a convolutional neural network (CNN) backbone and a graph attention network (GAT) backbone to extract discriminative features from ECG time-frequency spectrums and fNIRS prefrontal cortex (PFC) network respectively with deeper neuroscience domain knowledge, and eventually achieved 0.90 AUC. Results supported that handcrafted features extracted by specialized neuroscience methods can alleviate overfitting. Inspired by the small-world nature of the brain network, the fNIRS PFC network was organized as an undirected graph and embedded by GAT. It is proven to perform better in information aggregation and delivery compared to a simple non-linear transformation. The model provides a potential neuroergonomics application for evaluating the human state in vital human-cybernetics systems under industry 5.0 scenarios.


\end{abstract}

\bgroup
\color{abs}
\hrule
\egroup

%% file: files/02introduction.tex
\section{Introduction}

Industry 5.0 is centered around a new engineering approach that focuses on developing solutions that prioritize human needs \cite{indus5.0_XU2021530}. This emerging trend emphasizes the crucial role of human beings within systems \cite{Pilla2022}.
In this context, human-cybernetics systems require immediate and dependable comprehension of human states, encompassing both physical and mental aspects \cite{humanstatein5.0_10136627}. This is particularly crucial for evaluating operators' performance under intense thermal stress \cite{heat_9666525}, high mental load \cite{cogload_KARMAKAR2023104227}, heavy fatigue \cite{aei1MEHMOOD2022101777}, and severe drowsiness \cite{drowsiness_s21113786} in typical human-cybernetics systems, such as the air traffic control environment \cite{aei2LI2021101325, aei5XIONG2023101894}, industrial maintenance \cite{aei3SHI2020101153}, occupational training \cite{aei4LIU2020101048}
, construction safety \cite{cons_FANG2022101729}, and industrial emergency management \cite{emer_BUERKLE2021102179}. \par
Currently, neurophysiological monitoring has made significant development, including electroencephalogram (EEG), functional near-infrared spectroscopy (fNIRS), electrocardiogram (ECG), electrodermal activity (EDA), and electromyography (EMG). These bio-signals contain various neural and humoral information, reflecting fundamental aspects of the human state, including the brain's electrical activities, energy metabolism, neural activity, arousal level, and muscle fatigue respectively. The rapid development provides an emerging and effective technology for assessing the human state in an operational environment in real time \cite{phy1zhang_physiological_2023, phy2lohani_review_2019, phy3CHARLES2019221}. \par
As machine learning improves by leaps and bounds, enormous studies have investigated performance modeling utilizing neural networks based on psychological signals. Traditional supervised classification machine learning methods mainly include logistic regression, support vector machine (SVM), naive Bayesian (NB), decision tree (DT), random forest (RF), and adaboosting. These approaches are based on explicit mathematical models and rely on supervised labels. Sharma et al. \cite{56sharma_mental_2021} extracted sample entropy to obtain nonlinear EEG information and reduced dimension based on the Fisher score. The first 40 distinguishing features were selected to be trained by the Bayesian optimized K-nearest neighbor (Bo-KNN) algorithm and achieved a task load evaluation model with an accuracy of 96.07\%. Zhang et al. \cite{ress2ZHANG2020106287} evaluated expert-rated operator performance by an SVM, that fused eye tracking data, workload, situation awareness, skin conductance response, and respiratory features, which achieved a prediction accuracy of 75-83\%.  \par
With the advancement of computer memory and processing speed, deep learning networks with large parameters, built on the foundation of artificial neural networks (ANN), exhibit improved performance in feature learning and classification tasks. 
Tjolleng et al. \cite{61TJOLLENG2017326} extracted the time-frequency domain features of ECG under different cognitive loads and input them into a three-layer MLP network. By adjusting the number of neurons in the hidden layer, they found that the best performance was achieved with 15 neurons, ultimately reaching 82\% accuracy in the test dataset. Fan et al. \cite{ress1FAN2023109103} proposed a fNIRS-based ANN model to predict operators' performance both in distraction-situation and non-distraction situation for maritime safety. \par
The invention of the convolutional neural network (CNN) has greatly advanced deep learning in the area of image recognition, surpassing the multi-layer perceptron in efficiency.  It is well-suited for modeling the time-series signal \cite{21Dai_2020}, spectral maps \cite{64JIAO2018582}, time-frequency spectrums \cite{22Mao_2020}, and brain activation maps derived from physiological signals, often presented as images. Another feasible approach is to arrange handcrafted features in a timeline to create a matrix, as the input in a CNN \cite{63lim_driver_2016}. \par
The small-world nature of human brains has been widely investigated in the field of cognitive neuroscience, especially in disease diagnosis \cite{119844754}, brain-computer interface (BCI) \cite{12zafar_hybrid_2023}, and mental load classification \cite{65mazher_beyond_2022}. Therefore, graph neural networks (GNNs) have become a highly promising method to model neuro-imaging data, such as fNIRS \cite{17_10339984, 18_10003896} and EEG signals \cite{19_9871984, 20zhao_graph_2021}. Currently, enormous variations have emerged from GNNs, such as graph convolutional networks (GCNs) \cite{13GCNkipf_semi-supervised_2016}, graph attention networks (GATs) \cite{GATvelickovic_graph_2017}, and graph isomorphic networks (GINs) \cite{14GINxu_how_2018}, all of which can capture graph-theoretic information from physiological data and enhance model accuracy.  \par
The physiological and psychological states of vital occupational operators in high-risk industrial facilities often change dynamically over time, particularly during emergencies. Therefore, capturing time-series information in the data is necessary for dynamic monitoring and warning about the abnormal states of operators. The recurrent neural network (RNN) and the long short-term memory network (LSTM) fulfill this requirement by incorporating input from both the previous and current states with learnable weights. Shayesteh et al. \cite{SHAYESTEH2023106019} combined 2D CNN (on EEG projection images), 1D CNN (on photoplethysmogram and EDA signal), and LSTM to assess the cognitive load of workers dynamically. \par

Various deep-learning modeling methods driven by physiological data have been thoroughly investigated. Numerous variants have been proposed with distinct advantages in modeling different types of physiological features. These include MLP for 1D data (various handcrafted features vectors), CNN for 2D image-based data (time-frequency spectrums, functional connectivity matrices, activation maps, handcrafted features in timeline), GNN for graph-based data (brain functional networks), and LSTM for time-dependent data (time series). However, there is still a relative lack of performance evaluation models based on multi-physiological sources which can effectively detect the abnormal states of essential operators in real time. Meantime, most existing models have utilized shallow handcrafted features from bio-signals and typical deep learning architectures, leaving room for further exploration in: 1) applying specific neuroscience analysis methods to incorporate domain knowledge from neuroscience for obtaining deeper and richer bio-feature information (e.g., brain networks from fNIRS and time-frequency spectrums from ECG, rather than simple brain activations from fNIRS and temporal features from ECG). 2) selecting appropriate advanced neural network variants that can fully utilize the nature of different physiological signals (e.g., applying GAT for brain networks instead of 1D CNN on primitive handcrafted features). Moreover, most studies have focused on EEG signals with high temporal resolution but sensitive to motion artifacts, while little attention has been paid to ECG signals and fNIRS signals with high ecological validity. \par

To address the above challenges, this work established a performance evaluation model driven by domain knowledge from neuroscience based on multi-physiological data (ECG-fNIRS) and explored the effective deep learning architecture to extract valid discriminative features. By processing ECG and fNIRS signals to obtain time-frequency spectrums and the prefrontal cortex network respectively, richer domain knowledge from neuroscience was integrated to achieve better evaluation results. The selected appropriate variants, such as spectrums embedded by CNN and networks embedded by GAT, have improved information delivery and discriminative feature extraction. Moreover, the model fused multi-physiological data with high ecological validity which provides a potential neuroergonomics application for evaluating the human state in vital human-cybernetics systems under industry 5.0 scenarios. 
\par

The rest of the paper is organized as follows. Section 2 presents the physiological data acquisition, preprocessing, handcraft-feature extraction, and the architecture with detailed parameters of the performance evaluation model proposed in this manuscript. In section 3, the classification performance of four models was compared. In section 4, the confusion matrix, Receiver Operating Characteristic (ROC) curves, Area Under Curve (AUC), and t-distributed stochastic neighbor embedding (t-SNE) dimensionality reduction visualization are displayed to demonstrate the capability of the proposed models to extract discriminative features. In section 5, the limitations and contributions are discussed.

%% file: files/03method.tex
\section{Methodology}
\begin{figure}[H]
    \centering
    \includegraphics[width=1\linewidth]{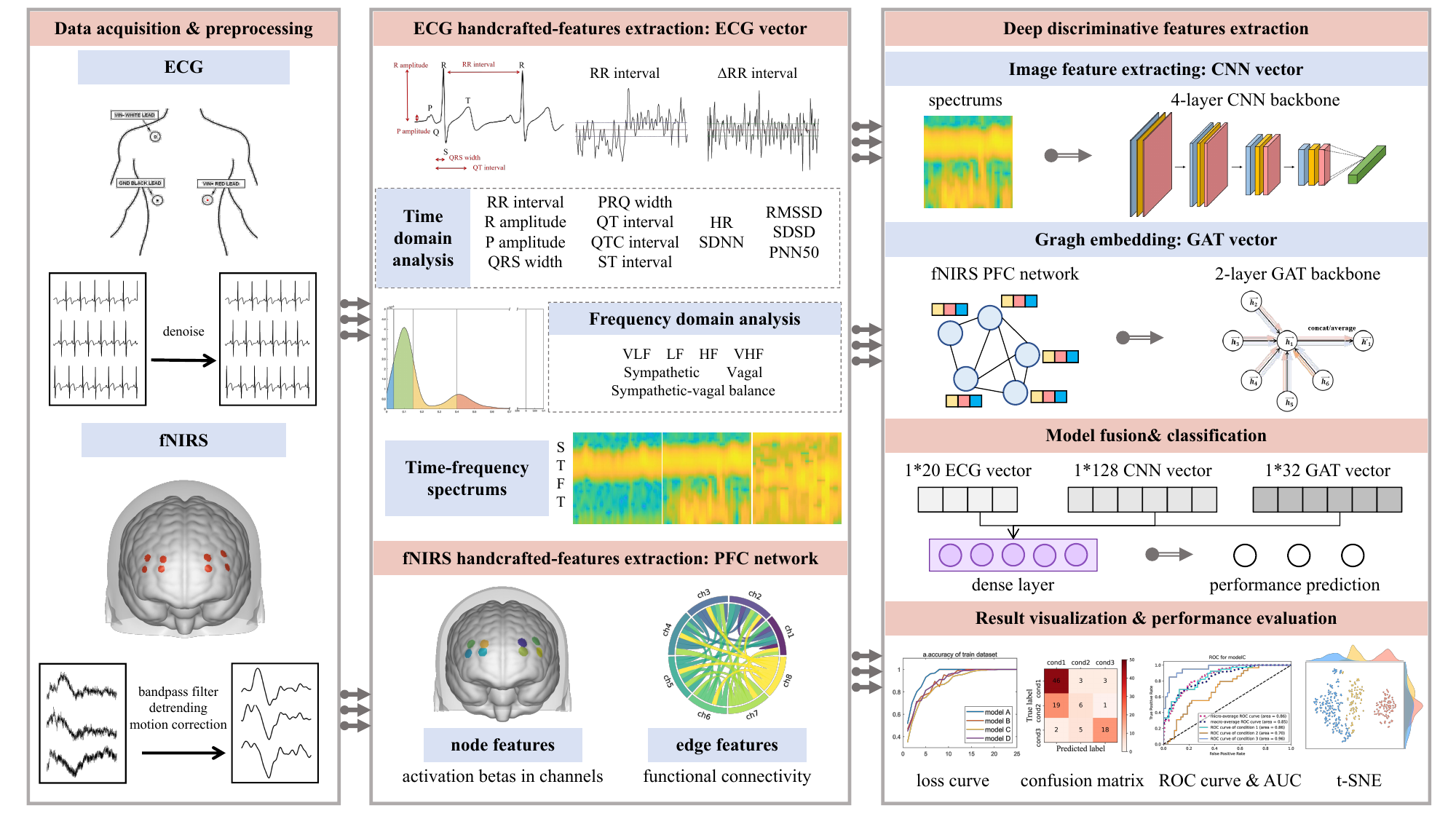}
    \caption{Overview of the processing flow in this paper. ECG time-frequency spectrums were obtained by STFT, and PFC networks were obtained with activation betas as node features and functional connectivity as edge features.}
    \label{fig:illustration}
\end{figure}

\Cref{fig:illustration} shows the outline of our performance evaluation model. The performance evaluation model consists of 5 parts: data acquisition \& preprocessing, ECG handcrafted-features extraction to obtain the ECG vector, fNIRS handcrafted-features extraction to obtain the prefrontal cortex (PFC) network, deep discriminative features extraction, and model fusion \& classification. Finally, various methods were applied to visualize the result and evaluate the performance, including the loss curve, ROC curve, confusion matrix, and t-SNE visualization. For ECG data, time-domain and frequency-domain analyses were applied to filtered ECG signals to extract ECG handcraft-feature vectors, including the amplitude and width of typical waves and power spectral density (PSD) of  typical frequency bands. Additionally, the short-time Fourier transform (STFT) was performed on ECG signals to obtain the time-frequency spectrums. Then, a spectrum-learned vector was extracted from each time-frequency spectrum through a 4-layer CNN backbone. For fNIRS data, generalized linear model (GLM) analysis and functional connectivity (FC) analysis were employed on filtered fNIRS signals to obtain betas (activation in brain regions) and FC strengths (synchronization among channels). fNIRS PFC network was formed with betas and FC strengths as node features and edge features respectively. The PFC network learned vector was extracted from each fNIRS PFC network through a 2-layer GAT backbone. Eventually, the performance category is predicted by a 3-classification dense layer based on a vector concatenating the ECG handcraft-feature vector, the spectrum-learned vector, and the PFC network learned vector. 

\subsection{Data acquisition, preprocessing, and handcraft-features extraction}
\subsubsection{data acquisition}
We implemented an experiment on operators of nuclear power plants (NPPs) (well-trained volunteers) under four hot-humid exposure scenarios (25℃ 60\% relative humidity, 30℃ 70\% relative humidity, 35℃ 80\% relative humidity, 40℃ 90\% relative humidity) to investigate the effect on their performance and typical cognitive and executive functions \cite{cog_exp}. To evaluate the general performance of operators in heat stress, the performance metrics selected in this experiment include performance in major NPPs tasks and essential cognitive functions:
\begin{itemize}
    \item \textbf{performance in major tasks}: According to NPPs operators’ cognitive model \cite{cogmodel1ohara2008human, cogmodel2whaley2016cognitive}, detecting and noticing, understanding and sensemaking, decision-making, action, and teamwork are the five essential cognitive and executive functions that operators require. The fault detection task and the reactor core shutdown operation task were conducted to encompass these essential functions. The faults involved the Steam Generator Tube Rupture (fault 1) and the rupture fault in the containment of the main water supply pipeline (fault 2). Participants should identify and verify the fault type and location following the emergency plan. In the operation task, they should confirm the key parameters and shut down the reactor following the emergency plan.
    \item \textbf{situation awareness (SA)}: SA reflects operators' perception, comprehension, and projection of the current state and future prediction of the critical system, which is highlighted by the Nuclear Regulatory Commission (NRC) \cite{SAWL} in the nuclear human factors engineering domain. SA is mainly assessed by the situation-awareness global assessment technique (SAGAT) \cite{sagat195097}.
    \item \textbf{workload (WL)}: workload encompasses operators' mental load and physical load, which is another factor emphasized by the Nuclear Regulatory Commission (NRC) \cite{SAWL} in the nuclear human factors engineering domain, typically showing a U-shaped correlation with performance. WL is mainly assessed by the NASA Task Load Index scale (NASA-TLX). 
    \item \textbf{working memory}: operators must handle enormous information flow and hold vital information in stressful situations, which is an important cognitive function known as working memory. Working memory is mainly assessed by the N-back task. 
\end{itemize}

\begin{figure}[H]
    \centering
    \includegraphics[width=1\linewidth]{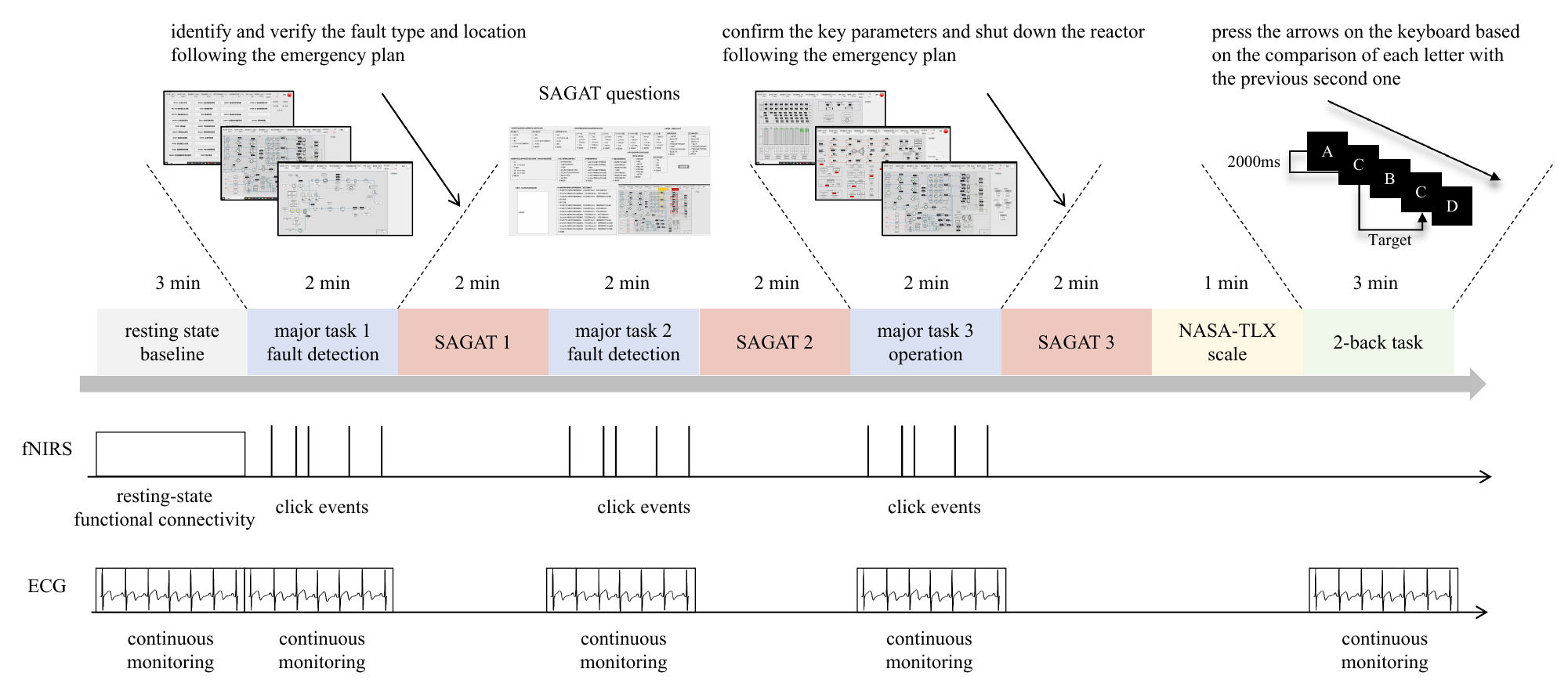}
    \caption{The experimental flow in each scenario.}
    \label{fig: exp flow}
\end{figure}

The experimental flow of each scenario is displayed in \cref{fig: exp flow}.  Every subject rested for 3 minutes to obtain the baseline and adapt to the environment at the beginning of a scenario. Then, they were asked to complete the major tasks and answer SAGAT questions alternately as soon as possible. Meanwhile, the duration, accuracy, reaction time, and error counts were recorded to assess the performance of major tasks. Next, they were required to answer specific questions related to the major tasks in SAGAT sections, and the accuracy and completion time were recorded to evaluate the capability of situation awareness. They then completed the NASA-TLX to indicate their workload level. Finally, they finished the 2-back task, a specific case of the N-back task, to assess their working memory based on reaction time, correct counts, error counts, and missed counts.
According to their performance metrics in different scenarios, their performance was divided into three categories following the conclusion of \cite{cog_exp} : 

\begin{itemize}
    \item category 1: good performance at a suitable workspace (25℃ and 60\% relative humidity \& 30℃ and 70\% relative humidity).
    \item category 2: decreased performance in a drowsiness state at a higher temperature and humidity environment ( 35℃ and 80\% relative humidity).
    \item category 3: a temporary slight improvement in simple reaction tasks but rapid impairment in advanced cognitive functions at the highest temperature and humidity. (40℃ and 90\% relative humidity)
\end{itemize}
    
Neural activity measured by ECG and energy metabolism measured by fNIRS were recorded in every scenario. Several exploratory approaches were applied to extract valid features that would be input data in the subsequent classification model training. \par
30 subjects (16 females and 14 males), aged 20-39 years (mean: 26±5 years) participated in the experiment.  Among them, four subjects' data were excluded due to incomplete recording. In addition, the data in scenario 3 of subject No.27 was missed because of an early recording interruption. Eventually, 515 pieces of ECG data and fNIRS data were obtained in total.

\subsubsection{ECG handcrafted features}
ECG signal was recorded by a 3-lead ECG module (BIOPAC MP160, BIOPAC Systems Inc., Santa Barbara, CA). \par
\textit{\textbf{Time-domain features}}. Signal filtering and QRS complex detection were used to extract the series of RR intervals. Various features were extracted based on the typical wave amplitudes, time width between different waves, and corresponding statistical magnitude.\par
\textit{\textbf{Frequency-domain features}}. In frequency-domain analysis, the fast Fourier transform (FFT) was performed on the filtered ECG data to calculate its power spectral density (PSD) in typical frequency bands: very low-frequency power (VLF) between [0, 0.04] Hz, low-frequency power (LF) between [0.04, 0.15] Hz, high-frequency power (HF) between [0.15, 0.4] Hz and very high-frequency power (VHF) between [0.4, 3]Hz \cite{ecgband1, ecgband2}. \par
ECG features in both the time and frequency domains are summarized in \Cref{table:ecg feature}. Thus, a 1*20 vector was extracted for each task completed by each subject in each scenario to represent neural activities based on ECG signals. 

\begin{table}[H]
    \centering
    \caption{ECG features extracted in time and frequency domain.}
    \begin{tabular}{p{1.5cm} p{3.4cm} p{1.2cm} p{9cm}}
        \toprule
        Domain  & ECG features & Unit  & Description   \\ 
        \midrule
        \multirow[t]{13}{1.5cm}{Time domain} & RR interval & [s] & Average of time distance between two adjacent R peaks  \\ 
        ~ & HR & [bpm] & Numbers of heartbeat every minute\\ 
        ~ & R amplitude & No unit & R wave peak amplitude   \\ 
        ~ & P amplitude & No unit & P wave peak amplitude  \\ 
        ~ & QRS width & [s] & Time between onset and end of the QRS complex  \\ 
        ~ & PRQ width & [s] & Time between onset of the P wave to the Q wave  \\ 
        ~ & QT interval & [s] & Time between the beginning of the Q wave and the end of the T wave  \\ 
        ~ & QTC interval & [s] & QT time interval divided by the square root of the RR interval  \\ 
        ~ & ST interval & [s] & Time between the S wave to the end of the T wave  \\ 
        ~ & SDNN & [s] & Standard deviation of RR time series    \\ 
        ~ & RMSSD & [s] & Root mean square of the difference of all subsequent RR intervals  \\ 
        ~ & SDSD & [s] & Standard deviation of the difference of all subsequent RR intervals  \\ 
        ~ & PNN50 & \% & Percentage of RR intervals in which the change of successive NN exceeds 50 ms  \\ 
        \midrule
        \multirow[t]{7}{1.5cm}{Frequency domain} & VLF  & [$\mathrm{s}^2$] & Spectral power of the RR time series in the band [0; 0.04 Hz]  \\ 
        ~ & LF  & [$\mathrm{s}^2$] & Spectral power of the RR time series in the band [0.04; 0.15 Hz]  \\ 
        ~ & HF  & [$\mathrm{s}^2$] & Spectral power of the RR time series in the band [0.15; 0.4 Hz]  \\ 
        ~ & VHF  & [$\mathrm{s}^2$] & Spectral power of the RR time series in the band [0.4; 3 Hz]  \\ 
        ~ & Sympathetic-vagal balance & No unit & Ratio between LF and HF  \\ 
        ~ & Sympathetic & No unit & Ratio between LF and (VLF+LF+HF)  \\ 
        ~ & Vagal & No unit & Ratio between HF and (VLF+LF+HF) \\ 
        \bottomrule
    \end{tabular}
    \label{table:ecg feature}
\end{table}

\subsubsection{ECG Time-frequency spectrums}
\begin{figure}[H]
    \centering
    \includegraphics[width=0.8\linewidth]{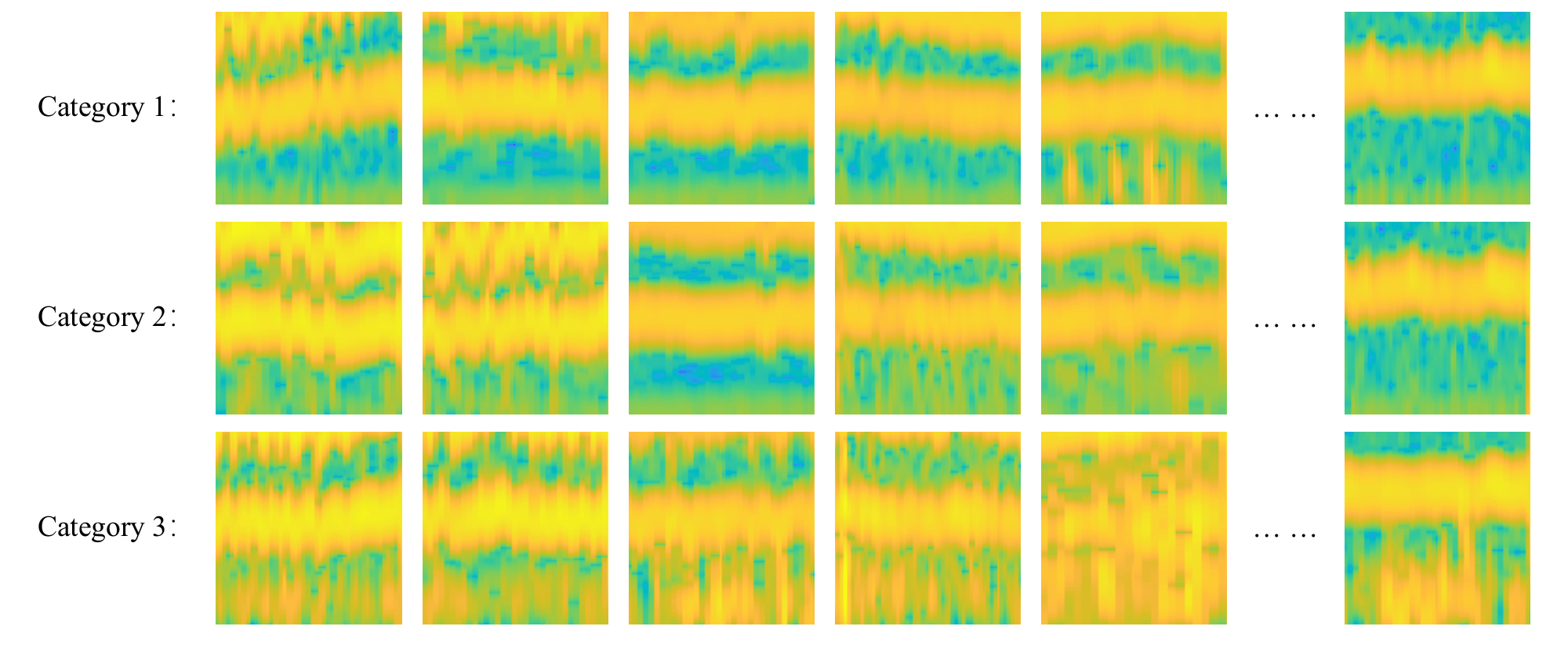}
    \caption{Some instances of the time-frequency spectrums resized to 64*64 pixels which are inputs to CNNs. Each row contains spectrums of different tasks from different subjects and is arranged in categories 1, 2, and 3 from top to bottom.}
    \label{fig:ecg tf plot}
\end{figure}

ECG data is a non-stationary signal. Therefore, STFT could be applied to ECG data to observe the change of frequency component and energy distribution over time. ECG data was scaled to [0, 100\%] in the timeline and STFT with [0, 3Hz] frequency limits, 0.5 frequency resolution, 20\% overlapping, and 0.875 leakage function was performed on the filtered ECG to get time-frequency spectrums. To obtain the input to CNNs, spectrums were resized to 64*64 pixels, and some instances of spectrums are plotted in \Cref{fig:ecg tf plot}.

\subsubsection{fNIRS features}
fNIRS data was measured by an 8-channel device (Artinis Octamon, Netherlands) placed on the prefrontal cortex (PFC). A bandpass filter ranging from 0.01Hz to 0.1Hz was arranged to denoise. First-order detrending was implemented to avoid baseline drift. Motion correction was applied to mitigate motion artifacts. \par
\textit{\textbf{Activated betas}}. After data preprocessing, betas were calculated by the generalized linear model (GLM). Beta values were derived by regressing the oxyhemoglobin (HBO) curve against the reference curve, which was generated by convolving the click event list with the hemodynamic response function (HRF). For input, betas were converted to a 1*24 vector (24 = 3 tasks * 8 channels).  \par
\textit{\textbf{Functional connectivity}}. For the resting-state fNIRS data, correlation coefficient, correlation coefficient after the fishier-Z transformation, coherence, and phase-locking value were calculated among channels to obtain functional connectivity (FC) matrix (8*8 symmetric matrix) and were flattened to a 1*112 vector (112 = 4 kinds of FC matrix * 28 channels combinations). \par

The correlation coefficient between the signal x and y is calculated by the following formula:
\begin{equation}
    R_{xy}=\frac{1}{N}\sum_{k=1}^{N}x(k)y(k)
\end{equation}

Coherence is calculated by the following formula, in which $S_{xy}(f)$ is cross-power spectral density between signal $x$ and $y$ at frequency $f$, $S_{xx}(f)$ and $S_{yy}(f)$ are auto-power spectral density for signal $x$ and $y$ at frequency $f$ respectively.
\begin{equation}
    \mathrm{COH}_{xy}=|K_{xy}(f)|^2=\frac{|S_{xy}(f)|^2}{S_{xx}(f)S_{yy}(f)}
\end{equation}

Phase-locking value is calculated by the following formula, in which $\Delta\phi_{rel}(t)$ is the phase difference between the complex signal $\Tilde{x}$ and $\Tilde{y}$ obtained by the Hilbert transform.
\begin{equation}
    \mathrm{PLV}=|\langle e^{i\Delta\phi_{rel}(t)}\rangle|=\bigg|\frac{1}{N}\sum_{n=1}^N e^{i\Delta\phi_{rel}(t_n)}\bigg|=\sqrt{\langle\cos\Delta\phi_{rel}(t)\rangle^2+\langle\sin\Delta\phi_{rel}(t)\rangle^2}
\end{equation}

Eventually, a 1*136 (136 = 24 + 112) vector for each subject in each scenario was extracted to represent energy metabolism from fNIRS data. NIRS-KIT toolbox \citep{nirskit} was utilized to assist in processing the fNIRS data.

\subsubsection{fNIRS PFC network}
\begin{figure}[H]
    \centering
    \includegraphics[width=0.9\linewidth]{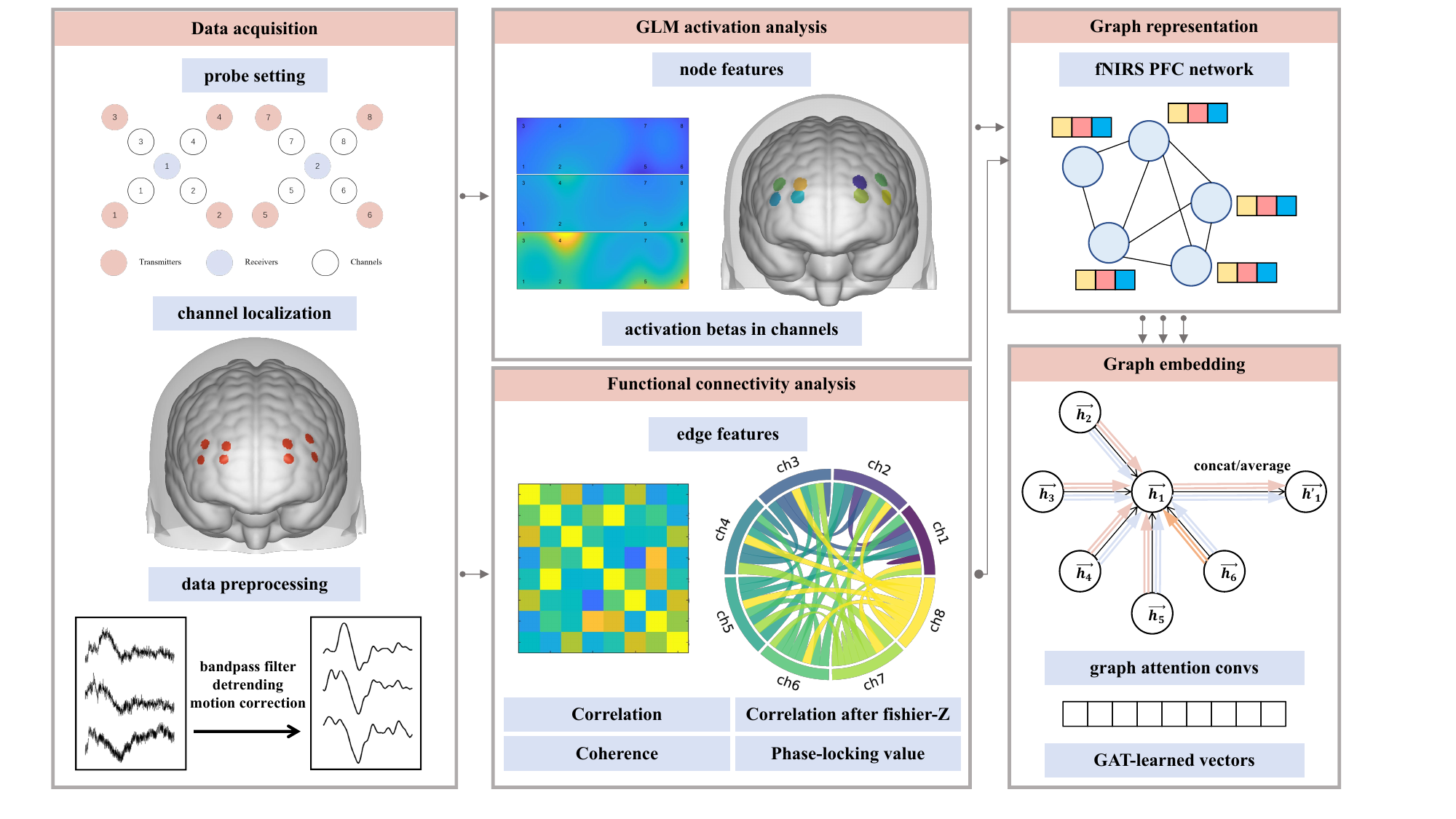}
    \caption{Illustration of fNIRS data process flow.} 
    \label{fig:PFCnet}
\end{figure}

Based on the small-world nature of the brain's functional network, the fNIRS signal was structured as the fNIRS PFC network, illustrated in \Cref{fig:PFCnet}. In the PFC network, fNIRS channels were nodes, the connections between channels were edges,  activated betas in fNIRS channels were node features, and functional connectivity between channels were edge features.

\subsection{Deep learning models and methods}
\begin{figure}[H]
    \centering
    \includegraphics[width=0.85\linewidth]{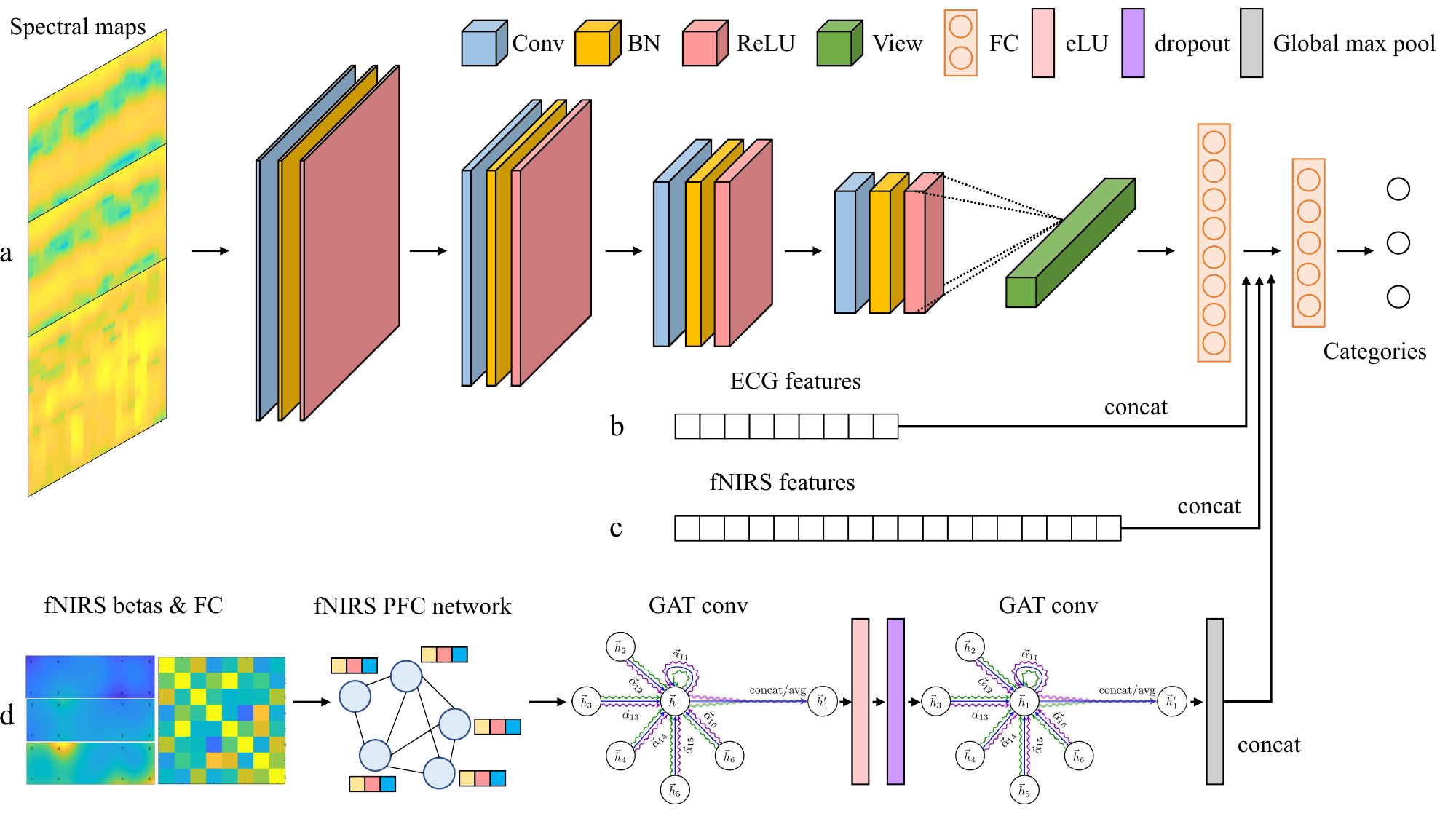}
    \caption{Illustration of our proposed models. Model A: a. model B: a+b. model C: a+b+c. model D: a+b+d. \textbf{a}: a three-classification model with a sequence of CNN convs based on spectral maps. \textbf{b}: 1*20 ECG vector was concatenated to the output features after the fully connected layer 1. \textbf{c}: 1*136 fNIRS vector was concatenated to the output features after the fully connected layer 1. \textbf{d}: 1*136 fNIRS vector was organized as a PFC network and embedded by GAT backbone to obtain GAT-learned vector. 1*32 GAT-learned vector was concatenated to the output features after the fully connected layer 1.}
    \label{fig:model struc}
\end{figure}

\begin{figure}[H]
    \centering
    \includegraphics[width=1\linewidth]{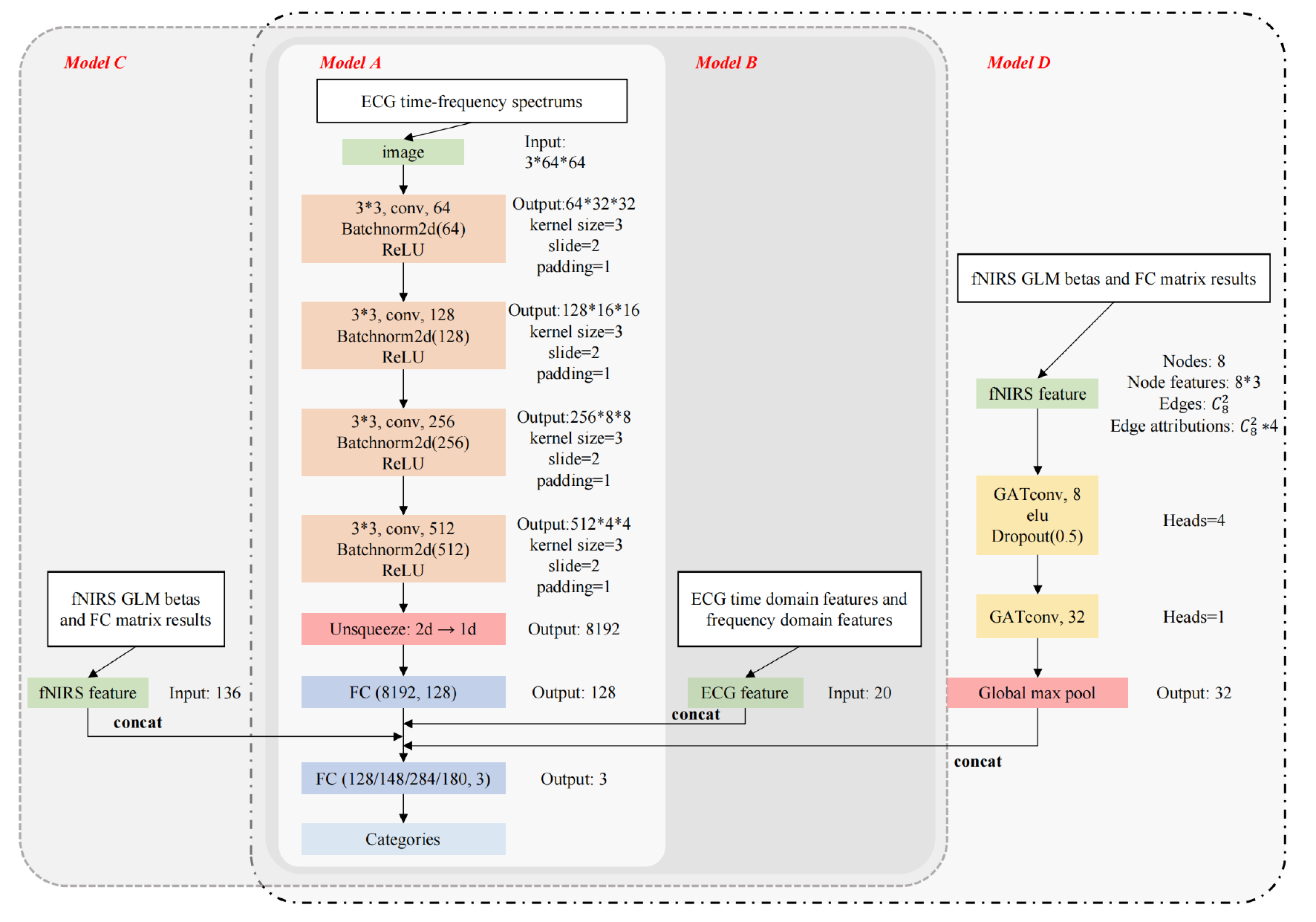}
    \caption{The architecture of four models with essential parameters.}
    \label{fig:model setting}
\end{figure}

To classify the performance in different scenarios using physiological data, CNNs were employed on time-frequency spectrums (model A). To improve the accuracy of the model, several models were developed to explore the effective architecture. The architectures of four models in detail are shown in \Cref{fig:model struc} and \Cref{fig:model setting}. 

\begin{itemize}
    \item Model A: CNNs on spectral maps.
    \item Model B: concatenate ECG features with CNN-learned features.
    \item Model C: concatenate ECG features and fNIRS features with CNN-learned features.
    \item Model D: concatenate ECG features and GAT-learned features from fNIRS with CNN-learned features.
\end{itemize}

\subsubsection{Graph attention networks (GATs)}
fNIRS can illustrate an executive functional network in the target brain area, with channels as nodes and the connections among channels as edges forming an undirected graph. Generally, the GLM is used in task-related data to calculate the activation of every channel that represents node features. Additionally, functional connectivities are obtained from resting-state data to calculate the synchronization among channels that represent edge attributions. This inspires us to apply graph neural networks (GNNs) to fNIRS features to extract effective information. \par
GAT \citep{GATvelickovic_graph_2017} is one important branch in GNNs presented by Petar in 2018. It brings in masked self-attentional layers to improve performance and decrease costs compared with conventional GNNs. It calculates hidden representations of each node by weighted summation from neighbor vectors. Each pair of neighboring nodes has a unique weight which is decided by the attention mechanism and reflects local structure. 

\subsubsection{Architecture of proposed models}
\textit{\textbf{Model A}}. In model A, the input image was resized to 64*64. Then, it went through 4 blocks composed of a sequence including a 3*3 convolution layer, batch-norm layer, and rectified linear unit (ReLU) function \cite{krizhevsky_imagenet_2017}. After that, the output feature was resized to 1 dimension and went through 2 fully connected layers to get predicted results. \par
\textit{\textbf{Model B}}. Based on model A, a 1*20 ECG handcraft-features vector was concatenated to the output of the fully connected layer 1 to provide more ECG temporal and spectral information in model B. Correspondingly, the number of input neurons in the fully connected layer 2 was adjusted to 148. \par
\textit{\textbf{Model C}}. Based on model B, a 1*136 fNIRS handcraft-features vector was concatenated to the output of the fully connected layer 1 to provide extra spatial fNIRS information in model C. Correspondingly, the number of input neurons in the fully connected layer 2 was adjusted to 284. \par
\textit{\textbf{Model D}}. Additionally, organizing fNIRS data into a graph and training it with GATs may yield better results by incorporating more structural information, as opposed to a basic nonlinear transformation following a fully connected layer on a 1-dimensional vector in model C. In model D, the 1*136 fNIRS handcraft-features vector was arranged as a graph with channels as nodes, GLM betas as node features, the connection between channels as edges, and functional connectivity as edge attributions. The graph feature was aggregated by a GAT convolution layer with 4 multi-head attention heads and 8 output channels. It was then activated by the exponential linear unit (eLU), which was effective in graph attention networks \cite{GATvelickovic_graph_2017}, and dropped out at 50\%. A subsequent GAT convolution layer and a global max pooling layer were used to derive the final representation of the input graph. A 1*32 vector learned by GATs and a 1*20 ECG handcraft-features vector were concatenated to the output of the fully connected layer 1. Correspondingly, the number of input neurons in fully connected layer 2 was adjusted to 180. \par
The main parameters of each layer in four models are mentioned in \cref{tab:model parameters} and the main parameters of the GATs module in model D are displayed in \cref{tab:gat parameters}. The input features of four models are summarized in \cref{tab:in_par}.

\begin{table}[H]
    \centering
    \caption{Main parameters of models. b reveals batch size.}
    \begin{tabular}{ccccccc}
        \toprule
         Name&  Input size&  Output size&  Kernel size&  Stride &  Padding & Numbers of neurons \\
        \midrule
         Conv1&  b*3*64*64&  b*64*32*32&  3&  2&  1& × \\
         Conv2&  b*64*32*32&  b*128*16*16&  3&  2&  1& × \\
         Conv3&  b*128*16*16&  b*256*32*32&  3&  2&  1& × \\
         Conv4&  b*256*8*8&  b*512*4*4&  3&  2&  1& × \\
         Resize&  b*512*4*4&  b*8192&  ×&  ×&  ×& × \\
         FC1&  b*8192&  b*128&  ×&  ×&  ×& 8192→128 \\
         FC2 in A&  b*128&  b*3&  ×&  ×&  ×& 128→3 \\
         FC2 in B&  b*148&  b*3&  ×&  ×&  ×& 148→3 \\
         FC2 in C&  b*284&  b*3&  ×&  ×&  ×& 284→3 \\
         FC2 in D& b*180& b*3& ×& ×& ×&180→3 \\
         \bottomrule
    \end{tabular}
    \label{tab:model parameters}
\end{table}

\begin{table}[H]
    \centering
    \caption{Main parameters of the GATs module in model D.}
    \begin{tabular}{ccccc}
        \toprule
         Names &  In channels&  Out channels&  Heads & Concatenation \\
         \midrule
         GAT conv1&  3&  8&  4& True \\
         GAT conv2&  32&  32&  1& False \\
         \bottomrule
    \end{tabular}
    \label{tab:gat parameters}
\end{table}

\begin{table}[H]
    \centering
    \caption{Input features of four models.}
    \begin{tabular}{p{0.8cm}p{4cm}p{10cm}}
        \toprule
         Model&  Input Features& Size\\
         \midrule
         A&  ECG time-frequency spectrums & 64 pixels * 64 pixels \\
         \midrule
         B&  ECG time-frequency spectrums & 64 pixels * 64 pixels \\
        &  ECG handcraft-features vector & 20 features: 13 time-domain features and 7 frequency-domain features\\
         \midrule
         C&  ECG time-frequency spectrums & 64 pixels * 64 pixels \\
        &  ECG handcraft-features vector & 20 features: 13 time-domain features and 7 frequency-domain features\\
        &  fNIRS handcraft-features vector& 136 features: 24 activated betas (3 tasks * 8 channels) and 112 functional connectivity strengths (4 kinds of FC matrix * 28 channels combinations)\\
         \midrule
         D&  ECG time-frequency spectrums & 64 pixels * 64 pixels \\
        &  ECG handcraft-features vector & 20 features: 13 time-domain features and 7 frequency-domain features\\
        &  PFC network & 8 nodes: 8 fNIRS channels \\
        & & 28 edges: $C_8^2$ fNIRS channels combination\\
        & & 24 node features: 24 activated betas (3 tasks * 8 channels) \\
        & & 112 edge features: 112 functional connectivity strengths (4 kinds of FC matrix * 28 channels combinations)\\
        \bottomrule
    \end{tabular}
    \label{tab:in_par}
\end{table}

\subsubsection{Network training}

\textit{\textbf{Data normalization}}. To ensure the comparability of different characteristics, each feature was normalized on its dimension. For spectral maps, mean value and standard deviation were calculated for each RGB channel across every pixel. For ECG and fNIRS features, mean value and standard deviation were calculated among each feature. In addition, all the features were standardized by the Z-score normalization following the formula below while $x$ is input data, $\mu$ is the average value and $\sigma$ is the standard deviation.

\begin{equation}
    x^*=\frac{x-\mu}{\sigma}
\end{equation}

\textit{\textbf{Cross-validation and test}}. The dataset was split into a training dataset and a test dataset in a 0.8:0.2 ratio. The fivefold cross-validation was performed on the training dataset to obtain the best hyperparameters. Specifically, the data in the training dataset is split into 5 folds. In each round, the model is trained on the four folds and evaluated on the left one-fold, which is the validation dataset. This process is repeated five times and the best average performance on the validation dataset corresponds to the optimal hyperparameter configuration. After fine-tuning the optimal hyper-parameters, the best model was trained on the entire training dataset and tested on the test dataset.
\par

\textit{\textbf{Hyper-parameters tuning}}. The proposed model includes several crucial hyperparameters that directly impact its performance. Therefore, the optimal hyperparameters were selected through grid search and 5-fold cross-validation (CV), with the results displayed in \Cref{tab:hyper-para}. The hyper-parameters included:

\begin{itemize}
    \item batch size: the number of examples utilized in one iteration. The best value is 128.
    \item weight decay: A coefficient to avoid overfitting. The best value is 1E-3.
    \item CNN layers: The convolution layers utilized to extract features from ECG time-frequency spectrums. The best value is 4.
    \item CNN hidden dimensions: The hidden channel dimensions of CNN Conv1. The hidden channel dimensions of subsequent CNN convolutional layers are doubled in size. The best value is 64.
    \item FC1 output dimensions: The output dimensions of fully connected layer 1. The best value is 128.
    \item GAT layers: The graph attention convolution layers utilized to extract features from PFC networks. The best value is 2.
    \item GAT heads: The number of multiple attention heads in GAT Conv1. The best value is 4.
    \item GAT hidden dimensions: The output channel dimensions of GAT Conv1. The best value is 8.
\end{itemize}

\begin{table}[H]
    \centering
    \caption{The performance of 5-fold CV in the grid search of each hyper-parameter.}
    \begin{tabular}{ccc} 
        \toprule
         hyper-parameter&  value&  5-fold CV micro-AUC\\
         \midrule
         batch size&  64&  0.8189±0.0276\\
         &  \textbf{128}&  \textbf{0.8267±0.0301}\\
         &  256&  0.8184±0.0370\\
         weight decay&  1E-2&  0.8200±0.0266\\
         & \textbf{1E-3}&\textbf{0.8251±0.0360}\\
         & 1E-4&0.8154±0.0330\\
         CNN layers& 3&0.8240±0.0252\\
         & \textbf{4}&\textbf{0.8241±0.0289}\\
         & 5&0.8129±0.0360\\
         CNN hidden dimensions& 32&0.8114±0.0349\\
         & \textbf{64}&\textbf{0.8211±0.0286}\\
         &  128&  0.8202±0.0252\\
         FC1 output dimensions& 64&0.8132±0.0245\\
         & \textbf{128}&\textbf{0.8241±0.0289}\\
         & 256&0.8238±0.0308\\
         GAT layers& 1&0.8187±0.0338\\
         & \textbf{2}&\textbf{0.8212±0.0265}\\
         & 3&0.8177±0.0341\\
         GAT heads& 3&0.812±0.0268\\
         & \textbf{4}&\textbf{0.8241±0.0289}\\
         & 5&0.821±0.0323\\
         GAT hidden dimensions& 4&0.8213±0.0336\\
         & \textbf{8}&\textbf{0.8251±0.0360}\\
         & 16&0.8212±0.0315\\
        \bottomrule
    \end{tabular}
    \label{tab:hyper-para}
\end{table}

\textit{\textbf{SGD optimizer with learning rate decay strategy}}. To accelerate convergence, the stochastic gradient descent (SGD) scheduler with 0.9 momentum was applied. Training epochs were limited to 40 because the loss on the validation set reached the optimal value and began to increase after that point. The cross-entropy loss function was selected to assess classification results. Various training strategies were implemented to enhance classification performance. To achieve faster and more stable convergence to the optimal solution, the ReduceLROnPlateau learning rate decay strategy was implemented in PyTorch and the initial learning rate was set to 0.001. This strategy involves halving the learning rate when the loss on the valid set stops decreasing for 5 epochs. Furthermore, the strategy will not be executed within 5 epochs of the last execution to prevent a sharp decrease in the learning rate.

%% file: files/04results.tex
\section{Results}

The classification performance of four different models is listed in \Cref{tab:summary loss acc} including accuracy and AUC (area under curve) of ROC curves. Model D achieves the best accuracy (81.82\% compared to 69.32\%, 70.45\%, and 68.18\% in models A, B, and C respectively in the test dataset. \par
In general, model D obtains the best performance with the highest AUC in micro-average (0.90), macro-average (0.86), category 1 (0.89),  category 2 (0.72), and category 3 (0.95). Model B achieves the second-best performance with a micro-average AUC of 0.85, a macro-average AUC of 0.79, and specific AUC values of 0.84, 0.56, and 0.92 for categories 1, 2, and 3, respectively. Model C ranks third with a micro-average AUC of 0.84, a macro-average AUC of 0.75, and specific AUC values of 0.80, 0.51, and 0.91 for categories 1, 2, and 3, respectively. Model A has the lowest AUC in micro-average at 0.84, macro-average at 0.75, category 1 at 0.79, and category 3 at 0.90. However, it shows a slight improvement in category 2 with an AUC of 0.53. \par

\begin{figure}[H]
    \centering
    \includegraphics[width=1\linewidth]{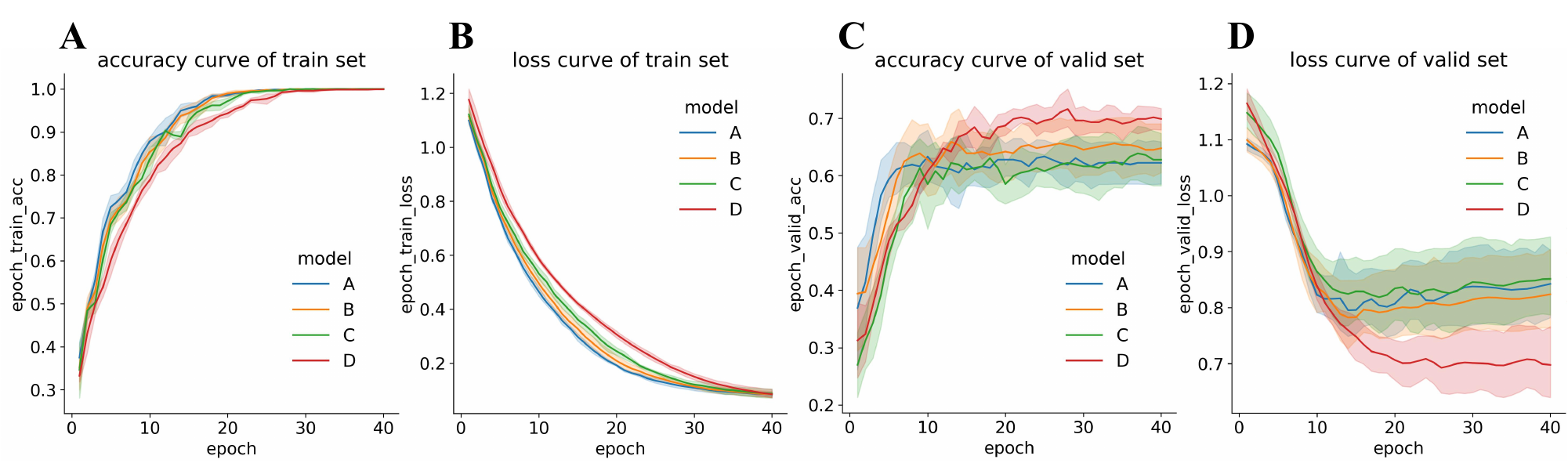}
    \caption{The learning curve of both the training dataset and validation dataset in 5-fold cross-validation. \textbf{A}: the accuracy curve for the training dataset. \textbf{B}: the loss curve for the training dataset. \textbf{C}: the accuracy curve for the validation dataset. \textbf{D}: the loss curve for the validation dataset.} 
    \label{fig:learning curve}
\end{figure}

In addition, the loss curves and the accuracy curves of four models in 5-fold cross-validation are plotted in \Cref{fig:learning curve}. Models A and B converge to the local optimal solution faster than models C and D in both the accuracy curve and loss curve. However, their loss in the validation dataset reaches a relatively low level and then increases after a few epochs, indicating overfitting. In contrast, loss in the validation dataset of models C and D still decreases and begins to increase at later epochs, which reveals a better capability of generalization. 
\par

\begin{table}[H]
    \centering
    \caption{Comparison of four models proposed. Including accuracy, loss, and AUC for the ROC curves of micro-average, macro-average, category 1, category 2, and category 3 on the test dataset. The bold values highlight the model with the best performance in each criterion.}
    \begin{tabular}{ccccccc}
        \toprule
         Model&  accuracy&  \multicolumn{5}{c}{AUC}\\
         &  &  micro-average&  macro-average&  category 1&  category 2& category 3\\
         \midrule
         A&  0.6932&  0.84&  0.75&  0.79&  0.53& 0.90\\
         B&  0.7045&  0.85&  0.79&  0.84&  0.56& 0.92\\
         C&  0.6818&  0.84&  0.75&  0.80&  0.51& 0.91\\
         D&  \textbf{0.8182}&  \textbf{0.90}&  \textbf{0.86}&  \textbf{0.89}&  \textbf{0.72}& \textbf{0.95}\\
         \bottomrule
    \end{tabular}
    \label{tab:summary loss acc}
\end{table}

%% file: files/05discussion.tex
\section{Discussion}
Based on the comparison of accuracy and AUC among various models, it is evident that spectral maps are efficient in classifying different performances in abnormal temperature and humidity scenarios. A 4-layer CNN (model A) has been proven to be an effective architecture with 69.32\% accuracy and 0.84 micro-average AUC.  \par
Furthermore, the superior AUC in models B and C compared to model A indicates that the handcrafted features, including ECG features and fNIRS features, can provide extra spatial, temporal, and spectral information. Despite the fact that the time-frequency spectrums of ECG signals have captured the frequency distribution over time, the size of 64*64 pixels was over-compressed to achieve a lighter network and faster classification speed. The simplest and most effective method to improve classification by fusing handcrafted features with ECG time-frequency spectrums is to concatenate the handcrafted features with the learned features through the 4-layer CNN directly from the spectrums. \par
However, model C shows worse classification despite more fNIRS data input than model B. This suggested that simple nonlinear transformations for manual fNIRS features are unable to extract effective information and deliver valid messages. Another reasonable explanation for the worse accuracy is that the CNNs-learned results were confused by the raw fNIRS manual features since a 1*136 fNIRS vector was too long compared with a 1*128 CNN-learned feature.  \par
Better performance in model D compared to model C reveals that the brain's functional network, as measured by fNIRS, exhibits characteristics of an undirected graph. Moreover, GNNs, in particular GATs, have been proven to be more effective tools for learning patterns of fNIRS features and brain networks. \par
As shown in plot d in \Cref{fig:learning curve}, the loss of models A and B decreases rapidly and then increases slowly after an early epoch point in the validation dataset, suggesting overfitting and reduced model generalization. The loss of models C and D reach the optimal point slower and increase after a later epoch, indicating that the employment of multi-physiological data can effectively alleviate overfitting compared to single-physiological data. \par

For a better understanding of the differences in classification performance for the four models, the confusion matrix, ROC curves, and t-SNE visualization are presented. \par
\begin{figure}[H]
    \centering
    \includegraphics[width=1\linewidth]{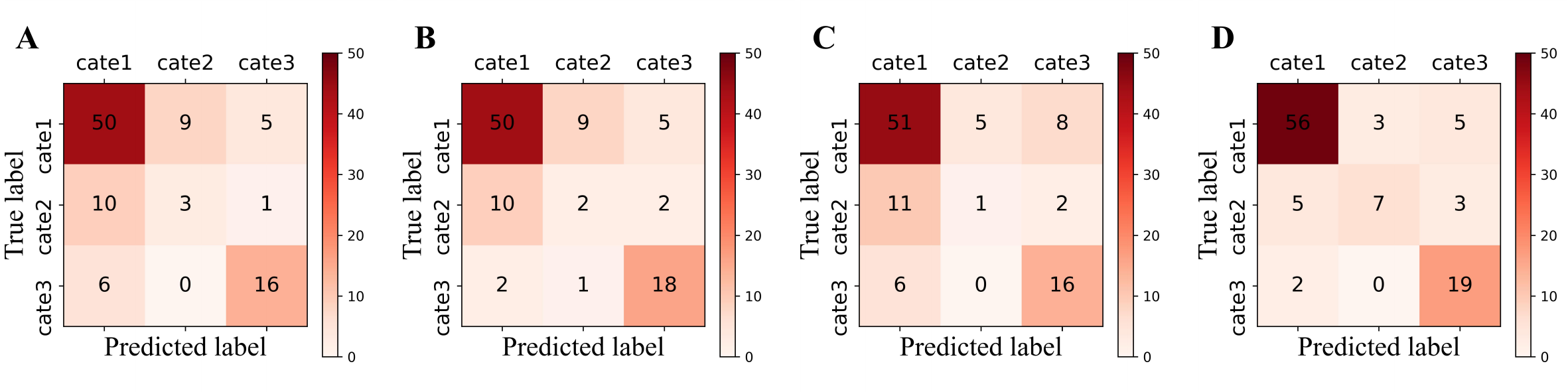}
    \caption{The confusion matrixes for models A, B, C, and D.}
    \label{fig:confuse matrix}
\end{figure}

The confusion matrixes for four models are presented in \Cref{fig:confuse matrix}. Models A and B incorrectly treat instances in category 2 and category 1, despite their good performance with category 1. Misclassification in category 1 and category 2 decreases significantly in models C and D compared to models A and B, especially for model D. Meanwhile, model D performs the best in classification for categories 1, 2, and 3. This result indicates a notable variance in the ECG signals between categories 1 and 3, while the pattern remains similar between categories 1 and 2. Nevertheless, fNIRS data can compensate for this drawback. However, all of them exhibit poor performance in classifying category 2. This could be attributed to the challenge of distinguishing physiological activities in non-extreme environments.

\begin{figure}[H]
    \centering
    \includegraphics[width=0.8\linewidth]{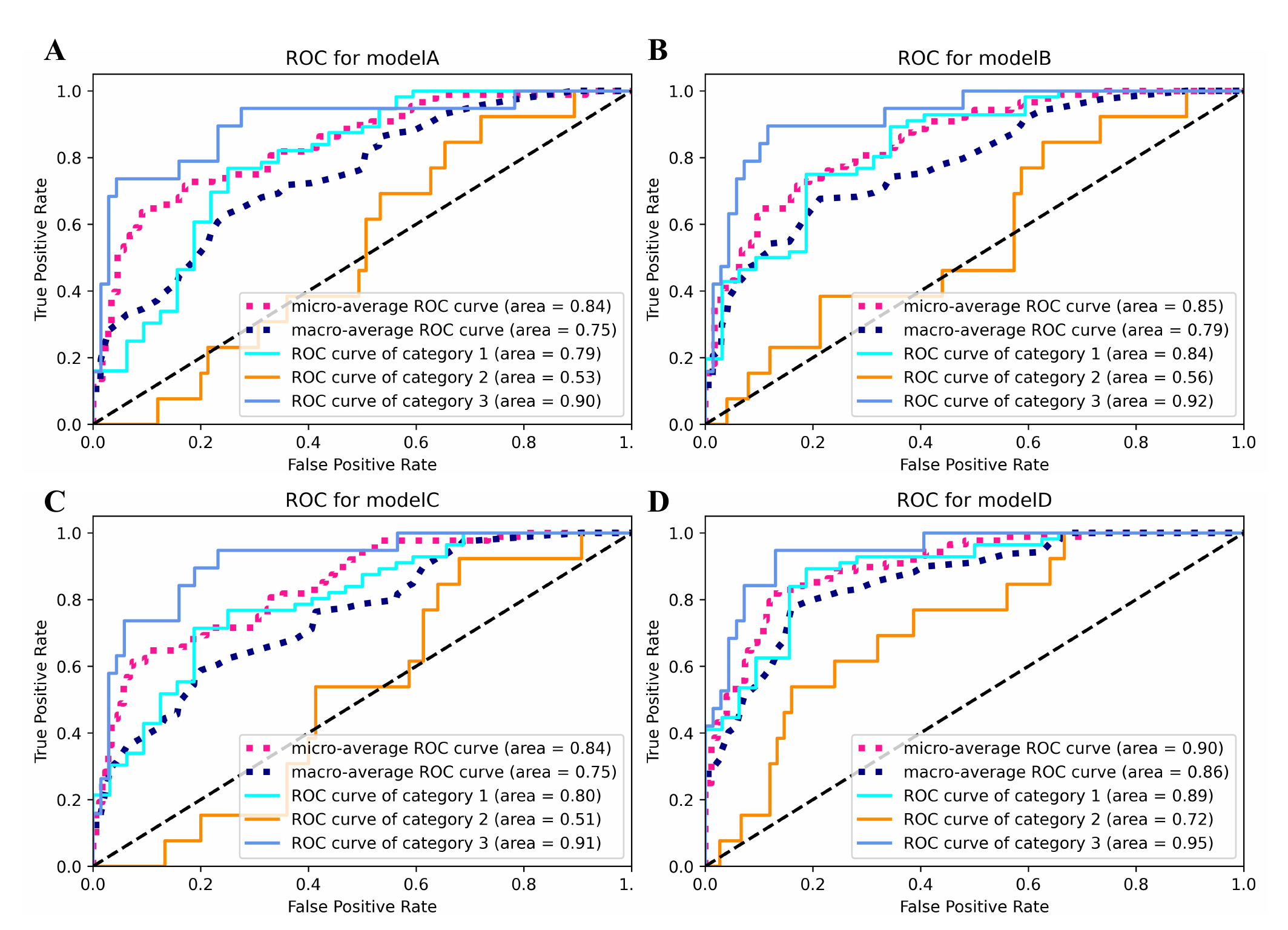}
    \caption{ROC curves for models A, B, C, and D, along with their corresponding micro-average, macro-average, category 1, category 2, and category 3. }
    \label{fig:roc}
\end{figure}

\begin{figure}[H]
    \centering
    \includegraphics[width=1\linewidth]{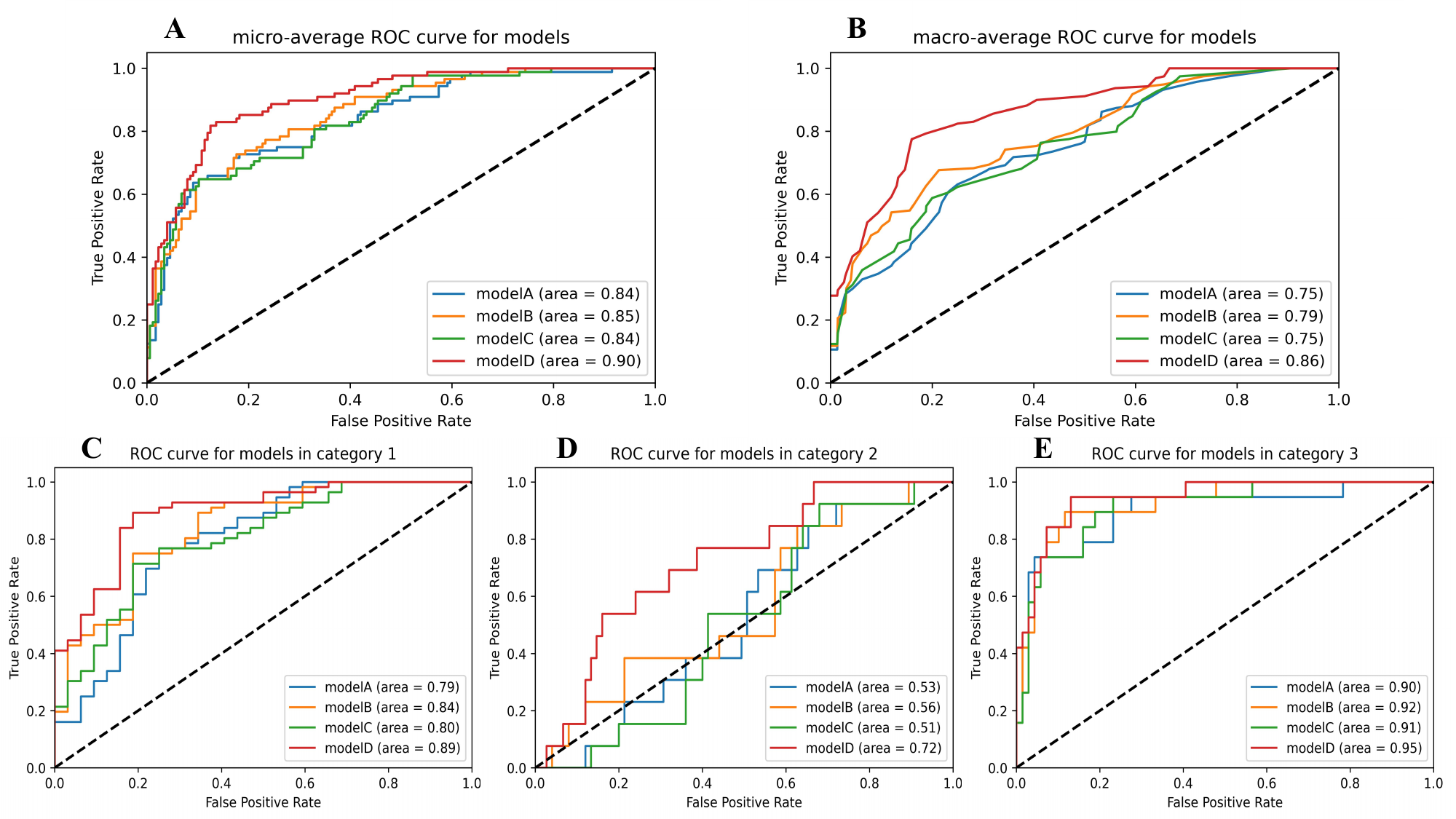}
    \caption{ROC curves for different criteria including micro-average (A), macro-average (B), category 1 (C), category 2 (D), and category 3 (E).}
    \label{fig:roc cond}
\end{figure}

ROC curves are displayed in \Cref{fig:roc} (for four models) and \Cref{fig:roc cond} (for five criteria). For all models, their best performance is in classification under category 3, followed by category 1, and worst in category 2, which is also supported by the confusion matrix. It reveals that it is competitive to monitor physiological data (i.e. ECG, fNIRS) to discover abnormal occupational states under extremely high temperature and humidity situations in real-time. \par
As shown in \cref{fig:roc cond}, model D exhibits an obvious advantage in performance evaluation. This result shows that organizing fNIRS data into an undirected graph (with activations as node features and functional connectivity as edge features) and embedding it through a 2-layer GAT network can effectively aggregate information, learn fNIRS patterns, and improve classification, which is also supported by the best accuracy and AUC in model D.

\begin{figure}[H]
    \centering
    \includegraphics[width=1\linewidth]{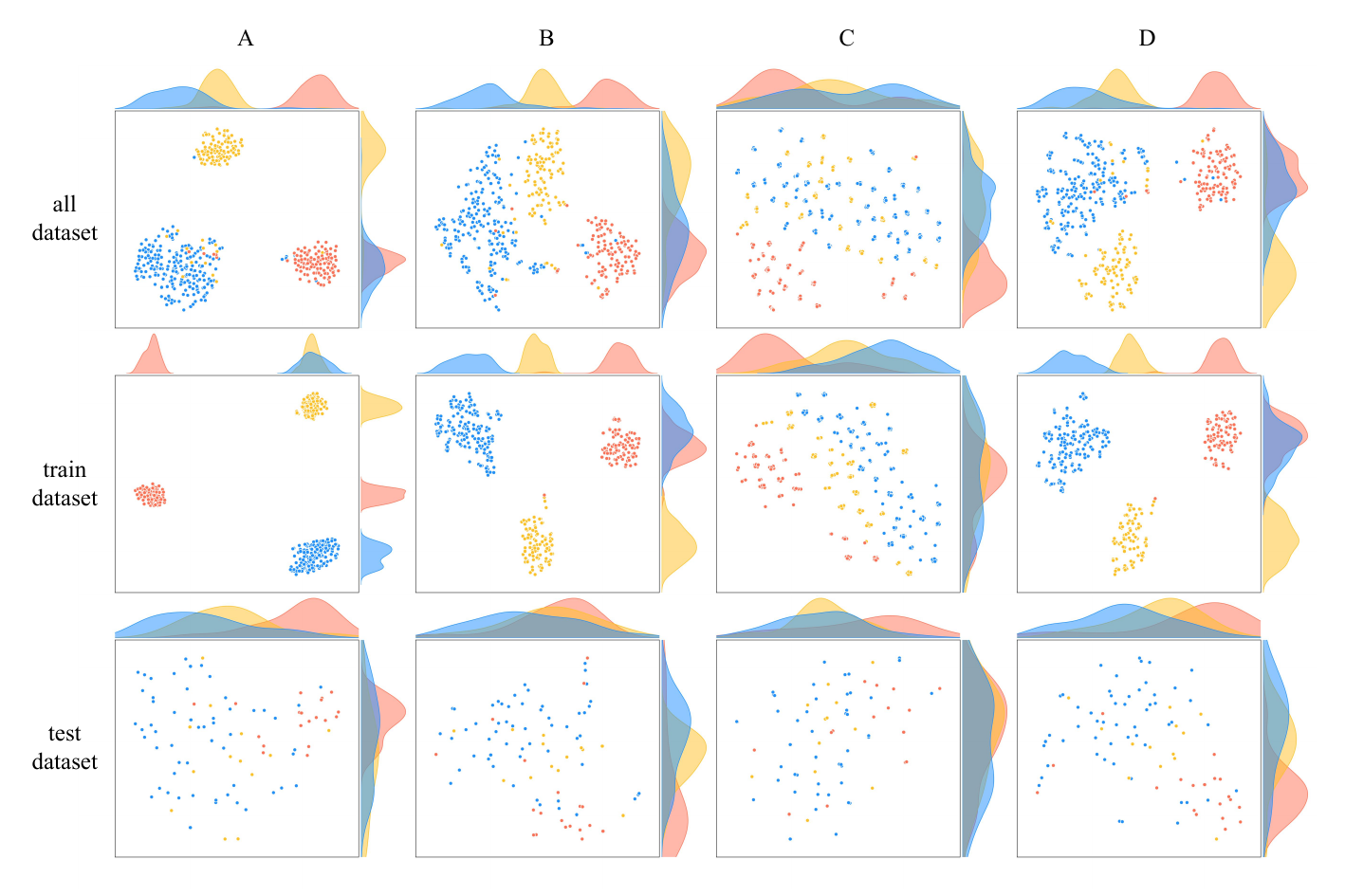}
    \caption{t-SNE plots for four models}
    \label{fig:tsne}
\end{figure}

To have a deep insight into the capability of deep models proposed in extracting discriminative features, t-distributed stochastic neighbor embedding (t-SNE) dimensionality reduction \citep{tsne} is conducted on the input vectors of the last fully-connected layer. This method embeds high-dimension vectors, usually with hundreds of dimensions and difficult to find hyperplanes to distinguish different classes directly, into a two or three-dimensional space, which is easy to visualize. In a t-SNE map, excellent classification is accompanied by tightly grouped clusters within the same class and obvious separation between different classes. Our t-SNE results are shown in \cref{fig:tsne}. For model A, t-SNE maps show that the learned features separate the most obvious in the training dataset but are mixed in the test dataset. This demonstrates that employing handcrafted features can alleviate overfitting in models B, C, and D. In model C, instances are always clustered in small groups, especially in the t-SNE map of the training dataset, because spectral maps were generated for every task in the experiment but they shared the same fNIRS data under the same scenario. This result shows that the poor classification in model C is mainly due to the excessive proportion of fNIRS data compared to ECG data. Therefore, GATs are adopted in model D to extract effective distinguished features from raw fNIRS data. A 136-dimension vector is reduced to a 32-dimension vector to achieve a balance between ECG data and fNIRS data. Furthermore, the most distinct peak in the probability density distribution of the test set in model D suggests that the fNIRS PFC network, going through the 2-layer GATs module, is effective in extracting discriminative features. However, the overlapping mode among categories shown in the probability density distribution, indicates that both fNIRS and ECG are unable to further separate category 1 and category 2, which is determined by the limitations of the physiological mechanism.

%% file: files/06conclusion.tex
\section{Conclusion}
In this paper, a CNN-GAT fused model is proposed for performance evaluation, driven by ECG time-frequency spectrums and fNIRS PFC networks, to identify abnormal occupational states in extreme temperature and humidity environments. A 4-layer CNN backbone (model A) processing time-frequency spectrums is proved to be practical, achieving a 69.32\% accuracy on a three-classification task. Furthermore, handcrafted ECG and fNIRS features are extracted and concatenated with learned features produced by CNNs. Unexpectedly, handcrafted features can effectively alleviate overfitting in models B and C, eventually achieving accuracy of 70.45\% and 68.18\% respectively. To enhance the learning effect of fNIRS data and incorporate domain knowledge from neuroscience, the advantage of the small-world nature of the brain network is taken into consideration, with channels as nodes, channel connections as edges, activation betas as node features, and functional connectivity as edge attributions. The PFC functional network is represented as an undirected graph and a 2-layer GAT backbone (model D) is applied to it, ultimately achieving 81.82\% accuracy and 0.90 AUC. The improved result shows that the GAT backbone obtained more discriminative features compared to a simple non-linear transformation on the handcrafted vector (model C). Additionally, the capability of each model to extract distinguishable deep features from original data is examined by ROC curves and t-SNE visualization, confirming the architectural validity of model D. Results reveal that the multi-physiological data (ECG-fNIRS), processed by a fused CNN-GAT backbone, effectively identify abnormal states in extreme environments (faster response but rapid impairment of advanced cognitive functions in category 3). However, it performs poorly in normal environments (distinguishing between category 1 and category 2), since the physiological mechanism is the primary constraint to further accuracy improvement.  \par

This work introduces graph attention networks into fNIRS data modeling. It proves that structuring fNIRS data as PFC networks and embedding it with GAT achieves better pattern-learning effects by leveraging the small-world nature of fNIRS and the advantages of GAT in graph embedding. Moreover, this work develops a multi-physiological fusion model for classifying performance under heat stress based on ECG signals and fNIRS signals, demonstrating high ecological validity. 
Additionally, this work provides a comprehensive framework and feasible model to detect the abnormal states of occupational operators by continuous physiological monitoring. This method overcomes the subjectivity of expert observation and the lag of post-testing, which provides a potential neuroergonomics application for evaluating the human state in vital human-cybernetics systems under industry 5.0 scenarios. \par

There is an abundance of future work to be pursued in this area. Due to the high cost of conducting precise experiments in the field of human factors, it is challenging to carry out large-scale experiments that involve hundreds or even thousands of subjects. Thus, there is still a relative lack of samples trained in the network, which restricts the model's generalization ability. To improve it, various data augmentation techniques could be applied, including simple methods like rotation, flip, and crop, as well as more advanced methods like generative adversarial networks (GANs).

%% file: files/07acknowledge.tex
\section{acknowledgment}
This study is supported by the State Key Laboratory of Nuclear Power Safety Technology and Equipment, China Nuclear Power Engineering Co., Ltd. (No.K-A2021.402), and the National Natural Science Foundation of China (No. 72374118, 72304165, 72204136).